\documentclass[preprint2]{aastex}

\newcommand {\ppmm}{P${}^2$M${}^2$}
\newcommand {\ppmmt}{P${}^2$M${}^2$~treecode}
\newcommand{\Vec}[1]{{\mbox{\boldmath{${#1}$}}}}
\newfont{\bg}{cmr10 scaled\magstep4}
\newcommand{\bigOl}{\smash{\hbox{\bg 0}}}
\newcommand{\bigOu}{\smash{\lower1.7ex\hbox{\bg 0}}}

\begin{document}

\title{Pseudoparticle Multipole Method: A Simple Method to Implement High-Accuracy Treecode}
\author{Atsushi Kawai}
\email{atsushi@atlas.riken.go.jp}
\affil{Computational Sciense Division, Advanced Computing Center, The
Institute of Physical and Chemical Research, 2-1 Hirosawa, Wako-shi,
Saitama 351-0198, Japan}
\and
\author{Junichiro Makino}
\affil{Department of Astronomy, School of Science, University of Tokyo,
7-3-1 Hongo, Bunkyo-ku, Tokyo 113-0033, Japan}

\begin{abstract}
In this letter we describe the pseudoparticle multipole method
(\ppmm), a new method to express multipole expansion by a distribution
of pseudoparticles. We can use this distribution of particles to
calculate high order terms in both the Barnes-Hut treecode and FMM.
The primary advantage of \ppmm~is that it works on GRAPE. GRAPE is a
special-purpose hardware for the calculation of gravitational force
between particles.  Although the treecode has been implemented on
GRAPE, we could handle terms only up to dipole, since GRAPE can
calculate forces from point-mass particles only. Thus the calculation
cost grows quickly when high accuracy is required. With \ppmm, the
multipole expansion is expressed by particles, and thus GRAPE can
calculate high order terms.  Using \ppmm, we implemented an
arbitrary-order treecode on GRAPE-4. Timing result shows GRAPE-4
accelerates the calculation by a factor between 10 (for low accuracy)
to 150 (for high accuracy). Even on general-purpose programmable
computers, our method offers the advantage that the mathematical
formulae and therefore the actual program is much simpler than that of
the direct implementation of multipole expansion.
\end{abstract}

\section{Introduction}

The calculation of the gravitational force is the most expensive part
of almost all $N$-body simulations.  The Barnes-Hut treecode
\citep{bh86} is a widely used algorithm to reduce the cost of the
force calculation. In the treecode, forces on a particles from distant
particles are replaced by multipole expansions of groups of particles. 
More distant particles are organized into larger groups, so that the
truncation error of the expansion is similar everywhere. Hierarchical
tree structure is used to form grouping efficiently. The calculation
cost is reduced from $O(N^2)$ to $O(N \log N)$.

Even with the treecode, the cost of force calculation is still high,
and it dominates the total calculation cost. In order to accelerate
the treecode further, we can use GRAPE \citep{s90,mt98}. GRAPE is a
special-purpose hardware for the calculation of gravitational force
between particles. It works in cooperation with a general-purpose
computer (host computer). The host computer does everything except for
the force calculation. The application of GRAPE to the treecode is
introduced by \citet{m91}, who implemented Barnes' (1990) modified
algorithm on GRAPE-1A \citep{f91}. \citet{a98} and \citet{k00}
reported its performance on GRAPE-3 and GRAPE-5, respectively. The
speedup factor they obtained is in the range of 30 to 50.

Although GRAPE can significantly accelerate the treecode, its
application has been limited to simulations where the requirement for
the accuracy is modest. Since GRAPE can only calculate forces from
point mass particles, we have not been able to handle terms of the
multipole expansion higher than dipole. Thus, the calculation cost
grows quickly when high accuracy is required.

In this letter, we introduce the pseudoparticle multipole method
(\ppmm) which makes it possible to evaluate higher-order expansions on
GRAPE. In \ppmm, multipole expansions are represented by a small
number of pseudoparticles. The masses and positions of pseudoparticles
are determined so that they have the same expansion coefficients as
the corresponding physical particles up to the specified order. With
the \ppmm, we can use GRAPE to evaluate high order terms, since they
are expressed by particles.

In this letter, we first describe the procedure to express 
quadrupole moment using three particles (\S~2). Then we briefly describe
the generalization to higher order expansion (\S~3). Finally we give the
result of numerical tests on GRAPE-4 (\S~4, \S~5).

\section{Quadrupole Moment with Pseudoparticles}

In \ppmm, we distribute the pseudoparticles so that they exactly
reproduce the coefficients of the multipole expansion of real
(physical) particles up to a given order. Conceptually, in order to
obtain such a distribution, first we calculate the expansion
coefficients from the distribution of physical particles. We then
solve the inverse problem to obtain the masses and positions of
pseudoparticles. In the following, we describe a practical procedure
to obtain the distribution of the pseudoparticles which can be used up
to quadrupoles.

In Cartesian coordinates, the multipole expansion up to quadrupoles of
the potential due to $N$ particles is expressed as
\begin{equation}\label{eq:mpicart}
\Phi(\Vec{r}) = \sum_{i=1}^N m_i \left\{ \frac{1}{r}
  + \frac{\Vec{r} \Vec{r_i}}{r^3}
  + \frac{1}{r^5}\left[\frac{3}{2}(\Vec{r} \Vec{r_i})^2-
    \frac{1}{2}r^2r_i^2 \right]
\right\}.
\end{equation}
The mass $M_j$ and the position $\Vec{R}_j$ of pseudoparticles must
be determined so that they give the same $\Phi$ up to a given order
$p$. In general, the expansion up to the $p$-th order has $(p+1)^2$
independent terms. 
Since each pseudoparticle has four degrees of
freedom (one for mass and three for position), theoretically we can
reproduce the expansion using
$K_{\rm min}(p) = \left\lceil (p+1)^2/4 \right\rceil$
pseudoparticles. Here $\lceil x \rceil$ denotes the minimum integer
not smaller than $x$.

\begin{figure}[htbp]
\includegraphics[width=80mm]{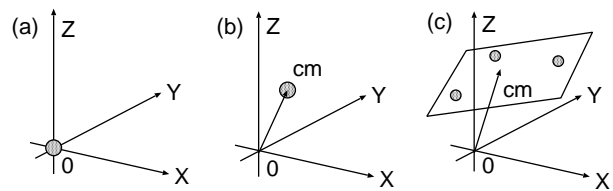}
\caption{The positions of pseudoparticles which reproduce
the multipole expansion up to monopole (left), dipole (middle), and
quadrupole (right). \label{fig:p012}}
\end{figure}

In order to express multipole expansion of order $p=0$, we need
$K_{\rm min}(0) = 1$ pseudoparticle. We must put the particle so that
$M_1$ and $\Vec{R}_1$ reproduce the first term (monopole term) of the
right hand side of equation (\ref{eq:mpicart}). For this purpose we
can set the mass $M_1 = M$, where $M$ is the total mass of physical
particles. At least formally, the position of the pseudoparticle can
be anywhere. In figure \ref{fig:p012} we place them at the origin as
example. In practice, we would never use zeroth order expansion since
we can achieve the first order accuracy with one particle, as we will
see below.

For $p=1$, $M_j$ and $\Vec{R}_j$ must reproduce the first and the
second term (dipole term) of the right hand side of equation
(\ref{eq:mpicart}). We can satisfy this condition by placing a single
pseudoparticle with mass $M$ at the center of mass of physical
particles, $\Vec{r}_{\rm cm}$ (see figure \ref{fig:p012}b), as is done with the
original Barnes-Hut treecode.

For $p=2$, the minimum number of pseudoparticles necessary is $K_{\rm
min}(2) = 3$. In the following, we'll see whether we can actually
construct the distribution of three pseudoparticles which reproduces
the multipole expansion up to quadrupole.

In order to reproduce the first and the second terms, the total mass
of the pseudoparticles should be $M$, and their center of mass should
be located at $\Vec{r}_{\rm cm}$. This is achieved by making their
total mass $M$ and center of mass to be the same as that of physical
particles.  In the following, we use a coordinate system shifted so
that $\Vec{r}_{\rm cm} = \Vec{0}$.

Pseudoparticles  should have the same quadrupole tensor
\begin{equation}
{\cal A} = \frac{3}{2} \sum_{i=1}^N m_i
\left[
\begin{array}{ccc}
{\displaystyle x_i x_i} &
{\displaystyle x_i y_i} &
{\displaystyle x_i z_i} \\
{\displaystyle y_i x_i} &
{\displaystyle y_i y_i} &
{\displaystyle y_i z_i} \\
{\displaystyle z_i x_i} &
{\displaystyle z_i y_i} &
{\displaystyle z_i z_i}
\end{array}
\right]
- {\displaystyle \frac{1}{2}\sum_{i=1}^N m_i \Vec{r}_i^2}
\end{equation}
as physical particles to reproduce the third term in equation
(\ref{eq:mpicart}). 

By definition, $\cal A$ is symmetric and traceless. Therefore we can
choose the coordinate axe so that $\cal A$ is diagonalized. In this
coordinate system, $\cal A$ is expressed as:
\begin{equation}\label{eq:qmat}
{\cal A} =
\left[
\begin{array}{ccc}
{\displaystyle a_1} & & \bigOu \\
& {\displaystyle a_2} & \\
\bigOl & & {\displaystyle -(a_1+a_2)}\\
\end{array}
\right].
\end{equation}
We can choose $a_1$ and $a_2$, so that the relation
\begin{equation}\label{eq:eigen}
  a_1 \ge a_2 \ge -(a_1+a_2)
\end{equation}
is satisfied.

Obviously, all three pseudoparticles should be on the x-y plane. 
Now our problem is reduced to determining the masses and positions of
three particles on the x-y plane so that they give the quadrupole
moment tensor of the form (\ref{eq:qmat}).

There are a variety of ways to satisfy this requirement. Here we give
just one example.

We still have three extra degrees of freedom, since we can change
masses and positions of two particles on the x-y plane freely, and we
have only three constraint. 
In our procedure, we set the masses as
$
  M_1 = M_2 = M_3 = M/3.
$
These masses reproduce the first term of the equation
(\ref{eq:mpicart}). Now we have only one extra degree of freedom
left. We use it by  setting $x$ component of
$\Vec{R_1}$ to 0. Now we can determine the position vectors as follows:
\begin{equation}\label{eq:p2s}
\Vec{R_1} = \left[
\begin{array}{c}
0\\
2 \beta \\
0
\end{array}
\right],
\Vec{R_2} = \left[
\begin{array}{c}
\alpha \\
-\beta \\
0
\end{array}
\right],
\Vec{R_3} = \left[
\begin{array}{c}
-\alpha \\
-\beta \\
0
\end{array}
\right],
\end{equation}
where $\alpha$ and $\beta$ are defined as
\begin{equation}
\alpha \equiv \sqrt{(2 a_1+a_2)/M},~~\beta \equiv \sqrt{(a_1+2 a_2)/(3M)}.
\end{equation}
Note that both $\alpha$ and $\beta$ are guaranteed to be real numbers. 
As we have already mentioned, this solution is not unique. For
example, if we set $y$, instead of $x$, component of $\Vec{R_1}$ to 0,
we obtain another solution that attains the same order of accuracy.

\section{Higher Order Generalization: Use of Spherical Design}

Here we describe the generalization of \ppmm~for multipole expansion
of higher order. Using the procedure we described in the previous
section, we can express multipole expansion up to quadrupole, with the
minimum number of pseudoparticles theoretically required. However,
the described procedure depends closely on the nature of
quadrupole-moment tensor, and it is difficult to extend this procedure
to higher orders.

\citet{m99} proposed a different approach based on the orthogonality
of spherical harmonics. His approach gives a systematic procedure to
obtain the distribution for an arbitrary order. Since his procedure
needs rather large number of pseudoparticles, it is not efficient when
the required accuracy is modest. But it may be useful for calculations
that require very high accuracy. In the following, we briefly describe
his procedure. Here we use spherical coordinates for mathematical
convenience.

The multipole expansion of the potential $\Phi(\Vec{r})$ is expressed
as
\begin{equation}\label{eq:mpe}
  \Phi(\Vec{r}) = \sum_{l=0}^\infty \sum_{m=-l}^{m=l}
                  \frac{\alpha_l^m}{r^{l+1}}Y_l^m(\theta, \phi)
\end{equation}
in spherical coordinates $\Vec{r} = (r, \theta, \phi)$.  Here
$Y_l^m(\theta,\phi)$ is the spherical harmonics and $\alpha_l^m$ are
the expansion coefficients. In order to approximate the potential
field due to the distribution of $N$ particles, the coefficients
should satisfy
\begin{equation}\label{eq:coeffi}
  \alpha_l^m = \frac{4\pi}{2l+1} \sum_{i=1}^N m_i r_i^l
               Y_l^{m*}(\theta_i, \phi_i),
\end{equation}
where $m_i$ and $\Vec{r}_i = (r_i, \theta_i, \phi_i)$ are the masses
and positions of the particles, and $*$ denotes the complex conjugate.
In order to express the expansion $\Phi(\Vec{r})$ up to the $p$-th
order, $K$ pseudoparticles must give the same coefficients
$\alpha_l^m$ for all $(p+1)^2$ combinations of $l$ and $m$ in the
range of $0 \le l \le p$ and $-l \le m \le l$. Thus, their masses
$M_j$ and the positions $\Vec{R}_j = (R_j, \theta_j, \phi_j)$ should
satisfy
\begin{equation}\label{eq:coeffij}
  \sum_{i=1}^N m_i r_i^l Y_l^{m*}(\theta_i, \phi_i) =
  \sum_{j=1}^K M_j R_j^l Y_l^{m*}(\theta_j, \phi_j),
\end{equation}
for $0 \le l \le p$ and $-l \le m \le l$. The mass $M_j$ and the
position $\Vec{R}_j$ are obtained as the solution of this equation.

As we have seen in the previous section, it is possible to directly
solve this equation for relatively small $p$, say, $p \le 2$. 
For large $p$, it is difficult because the equation is
nonlinear. In addition, it is not clear whether or not an acceptable
solution exists for arbitrary distribution of physical particles.

The key idea of Makino's approach is to fix the positions of
pseudoparticles. Equation (\ref{eq:coeffij}) is nonlinear for the
positions $\Vec{R}_j$, but is linear for the masses $M_j$. Thus, if we
fix the positions, the equation becomes linear. On the other hand, the
necessary number of pseudoparticles increases, since we can adjust
only the masses of pseudoparticles, and the degree of freedom assigned
to each particle is reduced from four to one.

We restrict the distribution of pseudoparticles to the surface of a
sphere centered at the origin, so that the the mathematics are further
simplified. With this restriction, the coefficients of multipole
expansion are expressed as the spherical harmonic expansion of the
masses of pseudoparticles. If we choose the positions of
pseudoparticles so that the numerical integration over their positions
retains the orthogonality of spherical harmonics up to the $p$-th
order, the mass of pseudoparticles are obtained as the inverse
transform of the expansion through numerical integration.

In Makino's procedure, the spherical $t$-design \citep{m63,hs96} is
used as the distribution of pseudoparticles. The spherical $t$-design
is defined as a set of points on a unit sphere with the following
characteristics. The summation of an arbitrary polynomial of a degree
at most $t$ over the points exactly match the integration over the
sphere (except for a constant coefficient).  Using spherical
$t$-design, we can achieve expansion order $p = \left\lfloor t/2
\right\rfloor$.  Here $\lfloor x \rfloor$ denotes the maximum integer
not larger than $x$.  For example, multipole expansion of order $p=1,
2, 3$ and $4$ are expressed with number of points $K=4, 12, 24$ and
$36$, respectively.  See \citet{hs96} for the number of points for
higher $p$, and the position coordinates of the points.

\section{Numerical Tests}
\label{sec:p2m2acc}

Using \ppmm, we implemented an arbitrary-order treecode (hereafter
\ppmmt) on GRAPE-4 \citep{m97,mt98} and a UNIX workstation. The source
code for these implementations are available upon request.  In the
following we show the results of numerical tests for the accuracy and
the performance. For the measurement, we used one GRAPE-4 processor
board (47 processors, 15 Gflops). The host computer for GRAPE-4 is a
COMPAQ AlphaStation XP1000 with Alpha 21264 processor, running at 500
MHz.

We uniformly distributed 262144 equal-mass particles within a unit
sphere. Then we calculated the force on each particle with \ppmmt, and
measured relative error $e$ and calculation time $T$ for various
values of the opening angle $\theta$. The error $e$ is defined as
r.m.s. relative difference from the exact force. As the exact force,
we used the force calculated with direct summation algorithm on the
host computer using IEEE-754 standard 64-bit arithmetic.

\begin{figure}[htbp]
\includegraphics[width=80mm]{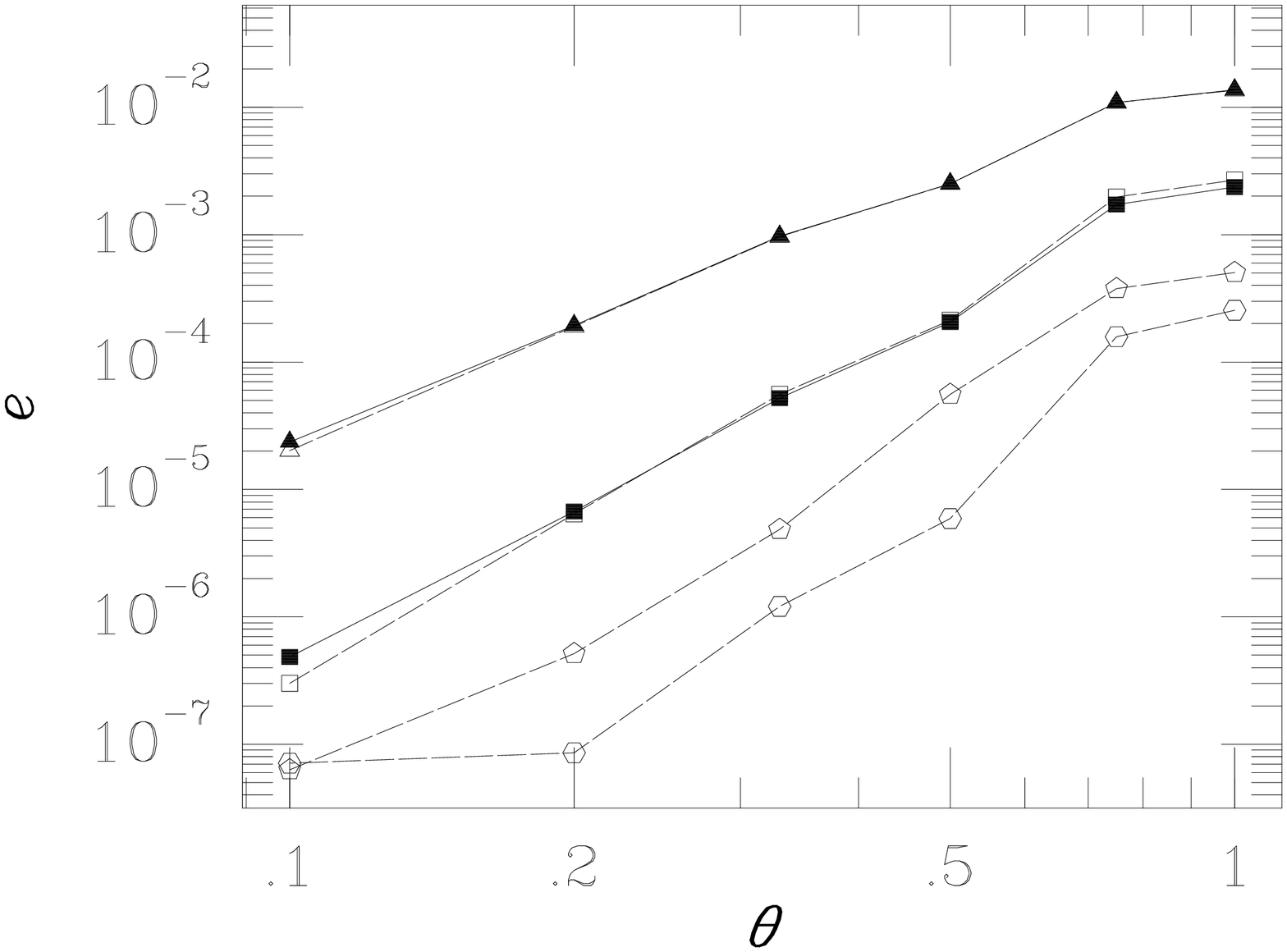}
\caption{The r.m.s. relative force error $e$ plotted against the
opening angle $\theta$. Solid curve with filled squares shows the
result of \ppmmt~of order $p=2$. Four long-dashed curves with open
triangle, square, pentagon, and hexagon are for generalized \ppmmt~of
$p=1, 2, 3$, and $4$, respectively. Solid curve with filled triangles
is for treecode of order $p=1$.
\label{fig:err}}
\end{figure}

\begin{figure*}[htbp]
\includegraphics[width=150mm]{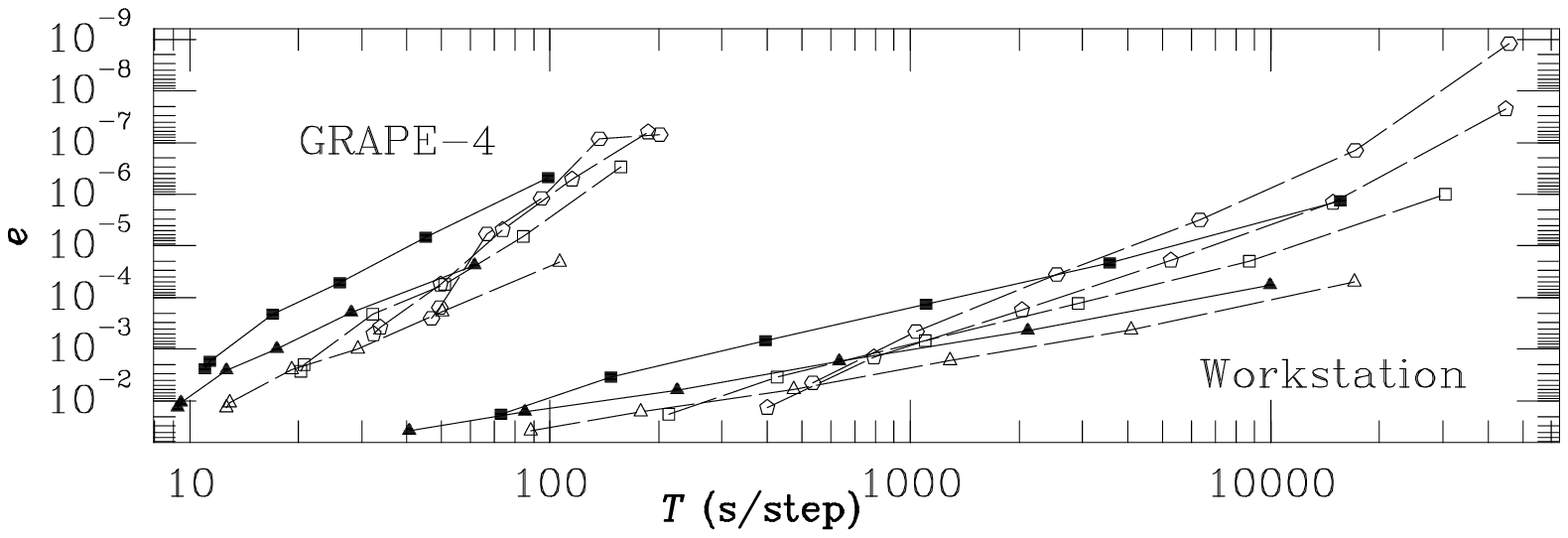}
\caption{The r.m.s. relative force error $e$ plotted against the
calculation time $T$. Meaning of the symbols are the same as those in
figure \protect{\ref{fig:err}}. \label{fig:time}}
\end{figure*}

Figures \ref{fig:err} and \ref{fig:time} show the results. In figure
\ref{fig:err}, the errors for \ppmmt~of orders $p=1$ through 4 are
plotted as functions of $\theta$. We can see that the error roughly
scales as $\theta^{p+1.5}$. This behavior agrees well with the
theoretical estimate given by \citet{m90}. The saturation of the
accuracy at $e \approx 10^{-7}$ is due to the hardware limitation of
GRAPE-4.

In figure \ref{fig:time}, the force error $e$ is plotted as functions
of the calculation time $T$. We can see that \ppmmt~of order $p=2$ is
the fastest for a wide range of accuracy ($10^{-7} \lesssim e \lesssim
10^{-3}$). For low accuracy ($e \gtrsim 10^{-3}$), the treecode of
order $p=1$ is the fastest. For extremely high accuracy ($e \lesssim
10^{-7}$), \ppmmt~of order $p=4$ is likely to be the fastest, if the
accuracy is not limited by the hardware.
We performed the same test without GRAPE-4, and found that GRAPE-4
accelerates the calculation by a factor of 10 (for $e \approx
10^{-2}$) to 150 (for $e \approx 10^{-6}$).

\section{An Example of  Simulation}
\label{sec:accdyng4}

Here we discuss the overall accuracy of the simulation with \ppmmt~on
GRAPE-4. We have already seen our code attain high accuracy in
calculation of instantaneous force, but this high accuracy does not
necessarily guarantee the accuracy of the total simulation. For
example, if the
force errors in consecutive timesteps  are strongly correlated, they
accumulate. As a result, overall accuracy of the orbits of stars might
be worse, compared to the case with weaker correlation.

We performed a simulation of the collision of two identical
galaxies. We chose the system of units so that the total mass of each
galaxy and the gravitational constant are both unity and the total
energy of each galaxy is $-1/4$. Galaxy model we used is the Hernquist
model \citep{h90} with 65536 equal-mass particles. We cut off the
distribution at radius 20.  We place the galaxies at initial
separation 3.0, and gave initial velocities so that they would follow
a parabolic head on collision. We integrated this system up to $t=16$
with time step $\Delta t = 1/200$ and softening parameter $\epsilon =
1/100$. We performed the same simulation using three different codes,
namely, \ppmm~treecode of order $p=2$, the treecode of order $p=1$,
and the direct summation algorithm.  For treecode simulations, the
opening angle $\theta$ is $0.5$.

\begin{figure}[htbp]
\includegraphics[width=80mm]{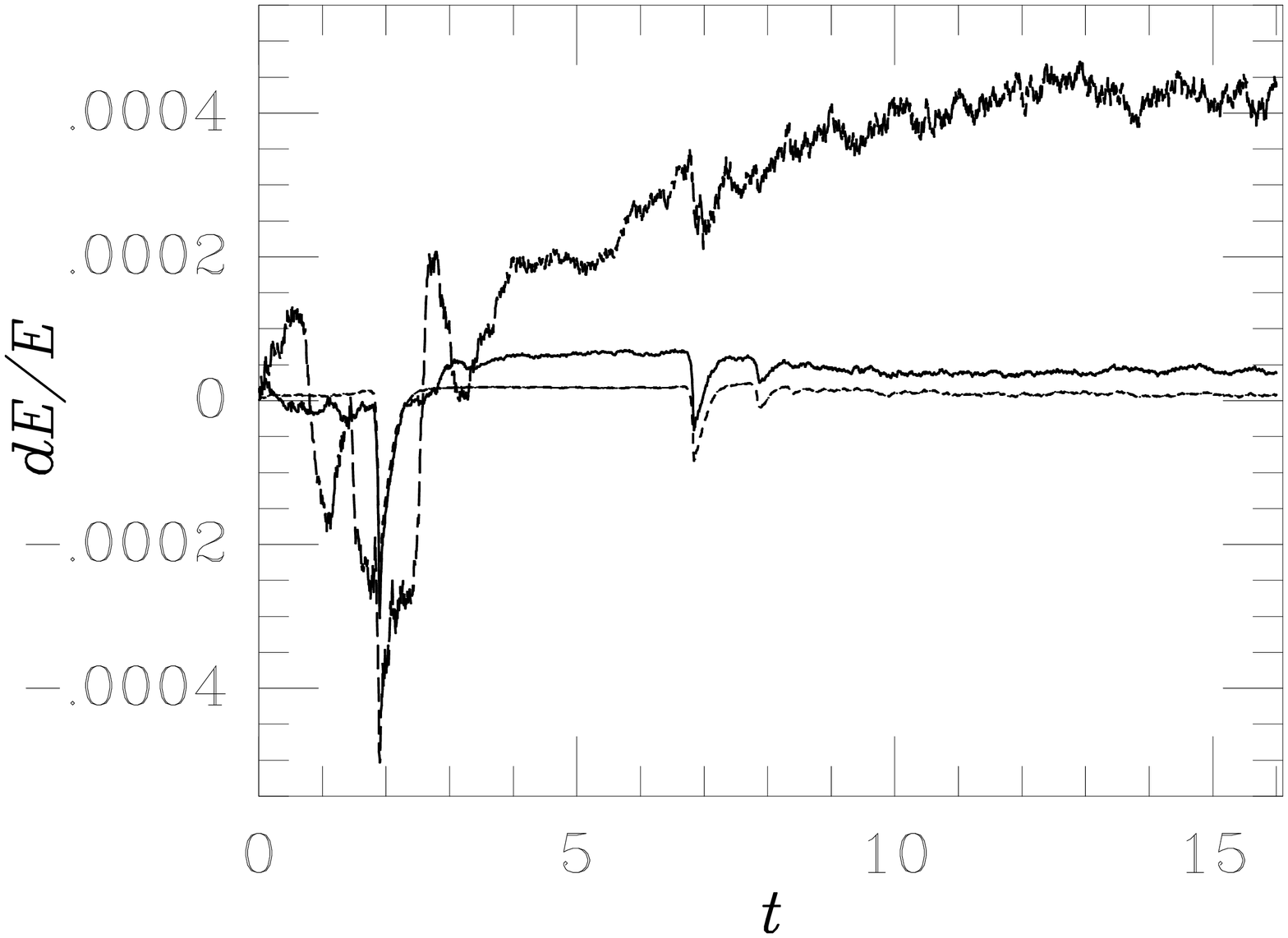}
\caption{The relative error of the total energy plotted
as a function of time. Solid, long-dashed and short-dashed curves are
for treecodes with $p=2$, $p=1$, and direct summation algorithm,
respectively.\label{fig:dyn}}
\end{figure}

Figure \ref{fig:dyn} shows the results. The relative error of the
total energy, $dE(t)/E(0)$, is plotted as a function of time $t$.  For
all runs, we can see three bumps around $t=2, 7$, and 8, which
correspond to close encounters of two galaxies. These bumps comes from
the time-integration error, since they can be seen in the result of
the direct summation algorithm with highly accurate force calculation. 
We can regard the deviation from the result of the direct summation as
being caused by the error of the force calculated with the treecodes.
For the treecode with $p=2$, the maximum deviation from the direct
summation is $5 \times 10^{-5}$, while for $p=1$ it is $4 \times
10^{-4}$. This a factor of 10 difference is consistent with the
difference of the accuracy of the instantaneous force shown in figure
\ref{fig:err}. Thus, we can conclude that the high accuracy in the
instantaneous force offered by our new \ppmm~does improve the overall
accuracy of the time integration.

\acknowledgements

We thank Piet Hut, Yoko Funato, Toshiyuki Fukushige and Toshikazu
Ebisuzaki for invaluable discussions concerning the subject of this
paper.  This work is supported  by the Research for the
Future Program of Japan Society for the Promotion of Science,
JSPS-RFTF 97P01102.

\end{document}